# Towards Maximum Optical Efficiency of Ensembles of Colloidal Nanorods

Owen Miller[1]*, Kyoungweon Park [2,3], Richard A. Vaia[2]*

[1]Department of Applied Physics and Energy Sciences Institute, Yale University, New Haven, Connecticut 06511, USA
[2] Materials and Manufacturing Directorate, Air Force Research Laboratory, Wright-Patterson AFB, Ohio 45433-7702, USA
[3]UES, Inc., Dayton, Ohio 45432, USA
*owen.miller@yale.edu, richard.vaia@us.af.mil

**Abstract**
Experimental and theoretical studies of colloidal nanoparticles have primarily focused on accurate characterization and simulation of observable characteristics, such as resonant wavelength. In this Letter, we tackle the *optimal design* of colloidal-nanoparticle ensembles: what is the largest possible optical response, which designs might achieve them, and can such response be experimentally demonstrated? We combine theory and experiment to answer each of these questions. We derive general bounds on the maximum cross-sections per volume, and we apply an analytical antenna model to show that resonant nanorods should nearly achieve such bounds. We use a modified seed-mediated synthesis approach to synthesize ensembles of gold nanorods with small polydispersity, i.e., small variations in size and aspect ratio. Polydispersity is the key determinant of how closely such ensembles can approach their respective bounds yet is difficult to characterize experimentally without near-field measurements. We show that a certain "extinction metric," connecting extinction cross-section per volume with the radiative efficiencies of the nanoparticles, offers a quantitative prediction of polydispersity via quantities that can be rapidly measured with far-field characterization tools. Our predictions apply generally across all plasmonic materials and offers a roadmap to the largest possible optical response of nanoparticle ensembles.

## I. Introduction

Colloidal nanorods combine strong plasmonic properties with high structural tunability, offering promise for applications ranging from optics to nanomedicine to obscurants.[1-5] For many applications, the key metric is the magnitude of the optical response relative to the volume or weight of the nanoparticles. Yet there have been significant challenges to fast and robust experimental characterization of such metrics, and there has been a gap in our theoretical understanding of the upper limits to such response. In this article, we theoretically predict the optimum efficiency of colloidal nanorods and experimentally demonstrate gold nanorods approaching their limits, combining a recently developed theoretical bound framework[6,7] with a modified seed-mediated nanoparticle synthesis approach and robust characterization techniques. We highlight the key role of reduced nanorod polydispersity (size and/or shape variations) in achieving the limits and develop a far-field scattering metric that provides rapid polydispersity characterization.

From a theoretical perspective, a continuum of techniques from analytical Mie-Gans and effective-medium theories to computational discretization and simulation has enabled modeling of individual or collective colloidal-nanoparticle properties.[8-11] Yet *optimal design* has been far less developed, due to the complexity of searching a combinatorially large design space. Recently, the advent of large-scale computational inverse design[12-18], as well as the development of theoretical frameworks for understanding bounds to light-matter interactions[6,7,19-26], have enabled preliminary works towards understanding optimal sizes, shapes, and material compositions. Yet there has been little work comparing fundamental limits to experimentally synthesized nanoparticles, and no identification of the key parameters that control whether such synthesized nanoparticles approach their respective bounds.

On the experimental side, there has been a significant lack of quantitative measurements of the magnitude of the optical response of colloidal nanorods. This stands in stark contrast to, for example, measurements of the resonant wavelengths of plasmonic nanoparticles, for which there are many careful measurements that match theoretical predictions.[9,27,28] Yet the contrast is not surprising given the many challenges to measuring the magnitude of optical response. First, to cover a wide bandwidth of interest, one must be able to synthesize a wide range of nanorod aspect ratios, as the aspect ratio is a primary determinant of the resonant frequency. Such coverage may require multiple synthesis protocols or highly controlled shape purity and byproduct creation, without which one may be limited to narrow ranges.[28] A second challenge is accurate structural characterization of the colloids. Conventionally, average size distribution data are obtained by image analysis of TEM images from a specific sub-population of nanoparticles. To obtain an unbiased size measurement, it is critical to select statistically representative images with sufficiently large sampling sizes. However, shape segregation and subjective sampling inherent to TEM sample preparation of polydisperse rod and rod-sphere mixtures tend to bias measurements . A third challenge is the measurement of molar extinction coefficients, for which the concentration of the constituent material, such as gold, requires measurements such as inductively coupled plasma-optical emission spectroscopy (ICP-OES), which is not readily available. Finally, whereas spectrophotometers are sufficient to measure extinction,[29,30] additional measurement data is needed to separately measure the absorption and scattering contributions to extinction. For gold nanorods, even just separate measurements of absorption and scattering have been limited to a small number of studies.[28]

In this article, we adapt theoretical approaches for identifying general bounds to light—matter interactions, developed in Refs. [6,7], to the specific problem of high-radiative-efficiency nanorods. We show that a simple optical-theorem-based radiative-efficiency constraint imposes bounds on the largest polarization currents that can be induced in the nanorods, resulting in bounds on how strong their response can be per unit volume of material (Sec. II). These bounds depend only on the optical susceptibility $\chi$ of the nanoparticles, the frequency $\omega$, and the radiative efficiency $\eta$, and are independent of the shape or size of the nanoparticles. Such bounds are inherently non-constructive, meaning it is not known whether they can be achieved with real structures. We apply an antenna-based circuit model[31] of subwavelength nanoparticles to show that properly designed nanorods can be expected to approach these global upper bounds through proper tuning of their sizes and aspect ratios (Sec. III). We validate the analytical antenna model with high-resolution boundary-element-method simulations[32,33] and optimizations. Motivated by these theoretical results, we describe a modified seed-mediated synthesis technique that enables high controllability of the aspect ratios of the nanorods (Sec. IV.) We describe a combination of approaches to quantitatively characterize the colloid optical response: statistical-bias reduction of the TEM-based structural-parameter measurements with scattering data, molar-extinction measurements enabled by combined ICP analysis and scattering data, and the use of both a spectrophotometer and an integrating sphere to independently measure absorption, scattering, and extinction. The synthesized nanoparticles reach within factors of 1.5–2.6 of the global bounds across visible and near-infrared frequencies. The measurements enable us to verify the theoretical prediction of polydispersity as the key constraint inhibiting the nanoparticles from reaching the bounds. From the antenna model, we identify a particular "extinction metric," which is the extinction cross-section per volume divided by $1 - \eta$, that appears well-suited for accurate characterization of the polydispersity of an ensemble of nanoparticles (Sec. V). This work lays a clear pathway to optimal design and synthesis of plasmonic nanoparticles, and we conclude with a perspective on new questions and opportunities prompted by these results (Sec. VI).

## II. Radiative-Efficiency Optical Cross-Section Bounds

In this section, we derive bounds to the largest possible cross-section of any plasmonic scatter, contingent on a minimum allowable radiative efficiency. In many plasmonics applications, large response (e.g. scattering or extinction) and high radiative efficiency (small optical absorption in the material) are competing objectives[34]. The highly subwavelength confinement associated with plasmonics originates in quasistatic resonances, which decouple the resonant wavelength from the size of the structure. Yet there is no radiation in quasistatic electromagnetism, and the only dissipation channel is material loss. To increase radiative loss the structure must become larger, but the introduction of radiation mitigates the quasistatic nature of the high confinement and tends to reduce the magnitude of the possible response. In this section, we quantify the tradeoff between large response and high radiative efficiency. We define a single optical figure of merit that captures this inherent tradeoff, given by σ$_{ext}$ / V / (1 – η), where σ$_{ext}$ / V is the extinction cross-section per scatterer volume, and η is the radiative efficiency.  As we discuss in the following sections, this quantity also serves as a far-field optical measure of the polydispersity of an ensemble of plasmonic nanoparticles.

We consider the maximum cross-sections that are possible for a given radiative efficiency. Optical cross-sections $\sigma$ are defined as the power scattered, absorbed, or extinguished from an incoming plane wave divided by the plane wave intensity, i.e., $\sigma_{\text{abs,scat,ext}} = \left(2Z_{\text{bg}}/|E_0|^2\right)P_{\text{abs,scat,ext}}$, where $E_0$ is the plane-wave amplitude and $Z_{\text{bg}}$ is the impedance of the background medium. Each of the powers $P_{\text{abs,scat,ext}}$ can be written in terms of work done by or on the polarization currents induced in the scatterer by the incident field. At a given frequency $\omega$, the polarization field $\boldsymbol{P}(x)$ is directly proportional to the electric field $\boldsymbol{E}(x)$ through its material susceptibility $\chi(\omega)$. Extinction is the work done by the incident field $\boldsymbol{E}_{\text{inc}}$ on the induced currents, $P_{\text{ext}} = (\omega/2) \operatorname{Im} \int_V \boldsymbol{E}_{\text{inc}}^* \cdot \boldsymbol{P} \, dx$ (known as the optical theorem[35,36]), absorption is the work done by the polarization currents on the total field, $P_{\text{abs}} = (\omega/2) \operatorname{Im} \int_V \boldsymbol{E}^* \cdot \boldsymbol{P} \, dx = (\omega \operatorname{Im} \chi(\omega)/2|\chi(\omega)|^2) \int_V |\boldsymbol{P}|^2 \, dx$, and the scattered power is the difference between the two. To derive bounds on the maximum cross-section for a given radiative efficiency, we extend the techniques of Refs. [6,7]. The key idea is as follows. The extinction power is *linear* in the polarization currents, while the absorbed power is a *quadratic* function of the polarization currents. Yet extinction must be larger than absorption, as scattered power must be nonnegative (in any passive system). This imposes a bound on how large the polarization field can be, simply by enforcing the constraint $P_{\text{abs}} \leq P_{\text{ext}}$. In this work, we extend this analysis by incorporating the radiative-efficiency constraint. Radiative efficiency is defined as the ratio of scattered power to extinction, $\eta_{\text{rad}} = P_{\text{scat}}/P_{\text{ext}}$. Requiring the radiative efficiency to be greater than or equal to some value $\eta$ implies that absorbed power must be smaller than the product of $1 - \eta$ with $P_{\text{ext}}$, leading to the constraint

$$P_{\text{abs}} \leq (1 - \eta) P_{\text{ext}}. \tag{1}$$

This constraint is a convex, quadratic constraint on the polarization field $\boldsymbol{P}$. To find upper bounds on the absorbed power, scattered power, or extinction, one can drop the more general constraint of Maxwell's equations, and only impose the constraint of Eq. (1). Then, through straightforward variational calculus, one can derive *analytical* upper bounds to the maximum cross-sections of any scatterer, a calculation performed in the SI. The resulting bounds, normalized to the volume $V$ of the scatterer, are

$$\frac{\sigma_{\text{ext}}}{V} \leq (1 - \eta) \frac{n_{\text{bg}} \omega}{c} \frac{|\chi|^2}{\operatorname{Im} \chi}, \tag{2}$$

$$\frac{\sigma_{\text{scat}}}{V} \leq \eta(1 - \eta) \frac{n_{\text{bg}} \omega}{c} \frac{|\chi|^2}{\operatorname{Im} \chi}, \tag{3}$$

$$\frac{\sigma_{\text{abs}}}{V} \leq (1 - \eta)^2 \frac{n_{\text{bg}} \omega}{c} \frac{|\chi|^2}{\operatorname{Im} \chi}, \tag{4}$$

where $n_{\text{bg}}$ is the refractive index of the background medium and $c$ is the speed of light. Eqs. (2-4) describe the maximum cross-sections per volume for scatterers of *any* shape. The primary mechanism to increase the possible cross-section is through the material "figure of merit"[6,37] $|\chi|^2/\operatorname{Im} \chi$, which describes the possibility to increase response through large magnitudes of the susceptibility, $|\chi|^2$, while the resonant response is inhibited by losses in proportion to $\operatorname{Im} \chi$. A natural question is whether the bounds of Eqs (2-4) are achievable. In the next section, we show through a coupled-mode circuit model that properly designed nanoparticles can approach these bounds, for many materials and radiative efficiencies.

## III. Antenna Model for High-Radiative-Efficiency Nanoparticles

The optical response of subwavelength plasmonic nanoparticles can be modeled with quantitative accuracy as lumped circuit elements[31,38-40]. Nanorods are typically dominated by electric-dipole resonances that can be treated with the introduction of a capacitance $C$ arising from the tip-to-tip charge separation, a small Faraday inductance $L_F$, a "kinetic" inductance $L_k$ (arising from nonzero electron mass) that typically dominates in plasmonic regimes, an Ohmic resistance $R_\Omega$ indicating the material absorption when currents are excited in the nanorods, and a radiation resistance $R_\text{rad}$ that encodes the power radiated to the far field by the same currents. Treating the nanoparticle as an *RLC* circuit, its resonant frequency $\omega_0$ is determined by the inductance and capacitance terms: $\omega_0 = 1/\sqrt{(L_F + L_k)C}$. This frequency $\omega_0$ is the frequency at which the absorption and scattering response of the electric-dipole mode of the nanoparticle will be maximum, and hereafter we consider the response at this resonant frequency.

To determine the cross-sections of a nanoparticle in this circuit model, we first need to model the incoming plane wave as a voltage source. An incoming plane wave carries an infinite amount of power, but there is a finite amount of power in the electric-dipole vector-spherical-wave, which is the only component that couples to the electric-dipole mode of the nanoparticle. This incoming power is given by $P_\text{inc} = \left(\frac{3\lambda^2}{2\pi}\right) I_\text{inc}$, where $I_\text{inc}$ is the intensity of the plane wave, i.e., $I_\text{inc} = |E_0|^2/2Z_\text{bg}$. To identify the available power, we equate the voltage across the radiation resistance of the nanoparticle to the incoming power,

$$\frac{V^2}{2R_\text{rad}} = P_\text{inc}, \tag{5}$$

which allows us to solve for the voltage, $V = \sqrt{2P_\text{inc}R_\text{rad}}$. The current excited in the nanoparticle is given by the voltage divided by the total resistance, $I = V/(R_\text{rad} + R_\Omega)$. The nanoparticle acts as a voltage divider circuit between the radiation and Ohmic resistances, and the scattering and absorption cross-sections are then proportional to the respective power delivery into these individual channels. Through a bit more algebra (cf. SI), we find that the cross-sections of the nanoparticles on resonance are given by

$$\sigma_\text{abs}(\omega_0) = \eta(1-\eta)\frac{3\lambda^2}{2\pi}, \tag{6}$$

$$\sigma_\text{scat}(\omega_0) = \eta^2 \frac{3\lambda^2}{2\pi}, \tag{7}$$

$$\sigma_\text{ext}(\omega_0) = \eta \frac{3\lambda^2}{2\pi}, \tag{8}$$

where $\lambda$ is the optical wavelength in the background medium and $\eta$ is the radiative efficiency, which in this circuit model is given by the ratio $\eta = R_\text{rad}/(R_\text{rad} + R_\Omega)$. Eqs (6-8) usefully predict that nanoparticle cross-sections divide the $3\lambda^2/2\pi$ relative power available to them amongst absorption and scattering, but they do not indicate any of the underlying dependencies on nanoparticle size or material.

The cross-section values of Eqs (6-8) can be connected to the susceptibility and volume of the nanorods through the radiative efficiency, $\eta = R_{\text{rad}}/(R_{\text{rad}} + R_\Omega)$. For an electric-dipole nanorod "antenna," the radiation and Ohmic resistances are given by the expressions[31]

$$R_{\text{rad}} = \frac{2\pi}{3} Z_{\text{bg}} \frac{\ell^2}{\lambda^2}, \tag{9}$$

$$R_\Omega = \frac{\lambda}{2\pi} Z_{\text{bg}} \frac{\text{Im}\,\chi}{|\chi|^2} \frac{\ell}{A} \tag{10}$$

where $\ell$ is the length of the nanorod and $A$ is its cross-sectional area. (The exact shape and curvature of the nanorods at their endcaps does not significantly affect these lumped-circuit-element parameters.) Inserting these resistances into the radiative efficiency expression gives:

$$\eta = \frac{V/\lambda^3}{V/\lambda^3 + \frac{3}{4\pi^2} \frac{\text{Im}\,\chi}{|\chi|^2}}. \tag{11}$$

Eq. (11) formalizes the intuition that radiative efficiency increases with scatterer volume (relative to cubic wavelength) and decreases with material losses (as measured by $\text{Im}\,\chi/|\chi|^2$). The quantitative accuracy of Eq (11) decreases in the extreme limits of radiative efficiency or material FOM, but as we show in Sec. V, it can be inverted to predict the volume of optimized nanorod designs quite accurately. Following straightforward algebra (SI) leads to the following expressions for the cross-sections per volume of nanorod particles:

$$\frac{\sigma_{\text{abs,np}}(\omega_0)}{V} = (1-\eta)^2 \frac{n_{\text{bg}}\omega}{c} \frac{|\chi|^2}{\text{Im}\,\chi}, \tag{12}$$

$$\frac{\sigma_{\text{scat,np}}(\omega_0)}{V} = \eta(1-\eta) \frac{n_{\text{bg}}\omega}{c} \frac{|\chi|^2}{\text{Im}\,\chi}, \tag{13}$$

$$\frac{\sigma_{\text{ext,np}}(\omega_0)}{V} = (1-\eta) \frac{n_{\text{bg}}\omega}{c} \frac{|\chi|^2}{\text{Im}\,\chi}. \tag{14}$$

Eqs (12-14), for the actual cross-sections of nanorods at their electric-dipole resonances, exactly equal the right-hand sides of Eqs. (2-4), which are the upper bounds on per-volume cross-sections of any scatterers with any shape and any multipolar response. This suggests that the bounds of Eqs. (2-4) should be approachable with properly designed nanorods.

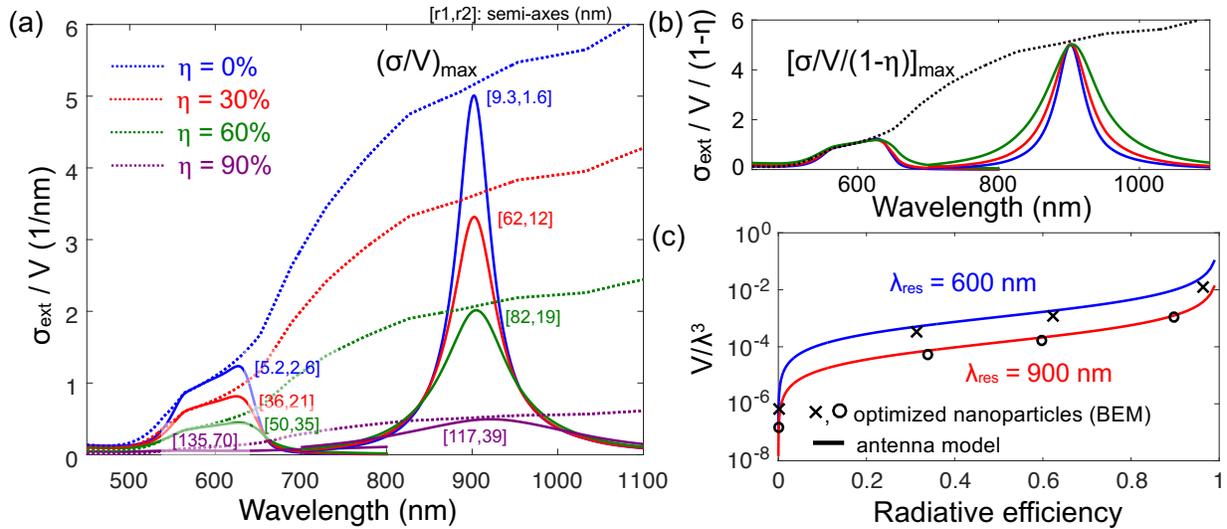

Figure 1. (a) Optimally designed gold nanorods, modeled as ellipsoids with semi-axis lengths in brackets, have extinction cross-sections per volume $\sigma_{ext}$ / V (solid lines, computed via BEM) that approach the analytical global bounds (dashed lines) of Eqs. (2-4). The maximum possible response decreases with increasing radiative efficiency $\eta$ (blue to purple). (b) The cross-section per volume divided by $1 - \eta$ collapses all bounds onto a single curve (black dashed), with the peaks of the simulated nanorod responses nearly coinciding. (c) Nanorods of a given material have two degrees of freedom, their aspect ratios and their volumes, which together determine their resonant wavelengths and radiative efficiencies. The antenna model relating these quantities together, in Eqs. (6-8) and Eqs. (12-14), predicts the solid curves in blue (600 nm wavelength) and red (900 nm wavelength). The BEM-simulated optimized nanorods have radiative efficiencies given by the markers ("x" for 600 nm wavelength, "o" for 900 nm wavelength), showing close agreement with the antenna model.

Figure 1 theoretically verifies that optimally designed gold nanorods can approach the bounds of Eqs (2-4). Fig. 1(a) shows the bounds on extinction cross-section per volume (dashed lines) for four minimum radiative efficiencies, ranging from 0% to 90%, with the bounds decreasing as minimum radiative efficiency increases. The solid lines in Fig. 1(a) are boundary element method (BEM) simulations of gold nanoparticles whose dimensions have been optimized with a free-software implementation[41] of the COBYLA gradient-free local-optimization algorithm[42]. One can see that the response of the nanoparticles, free of any circuit-model approximations, approach within 10% (and typically even closer) of the global bounds. The accuracy of the circuit-model theory is further confirmed in Fig. 1(c), where for both 600 nm and 900 nm wavelengths, the volumes of the optimal nanorods are shown as a function of radiative efficiency (markers), and the circuit-theory predictions (solid lines) run directly through the markers. Note from Eq. (11), and corroborated by this figure, that knowledge of the resonant wavelength and radiative efficiency is sufficient to identify the volumes of the individual nanorods as well. More broadly, properly designed nanorods should be globally optimal for absorbing, scattering, and extinguishing radiation.

## IV. Experimental results

In this section, we describe the experimental synthesis of nanoparticles that approach the optimal designs and bounds developed in the previous sections. We use a recently developed one-pot seed-mediated synthesis of gold nanorod (AuNR) ensembles. The synthesized ensembles have a range of average aspect ratios, from 2.15 to 7.69, with identical nucleation steps for each but variations in the timing and volume of the addition of a second growth solution (detailed methods in the SI). The resulting products are purified via centrifuge (at 3000 rpm for 10 min.) to remove excess CTAB and Ag complex. As opposed to conventional seed-mediated synthesis, modifications of the temporal control of the seed and reactant concentration for optimized seed development favor symmetry breaking towards rod formation and narrow polydispersity.[43] This method produces AuNRs centered at any of a wide range of aspect ratios, with narrow distributions of nanoparticle dimensions, aspect ratios and volume. While retaining a simple growth-solution composition (CTAB, $HAuCl_4$, $AgNO_3$, and hydroquinone), this method enables high shape purity (>96%), avoiding extraneous spheres and cubes that occupy volume but contribute almost zero extinction at longer wavelengths, where they are off-resonant.

Figure 2 shows representative optical measurements and STEM images of two example AuNR ensembles. Labeling the AuNR ensembles from 1 to 10 based on their aspect ratios (1 smallest, 10 largest), the two ensembles of Fig. 2(b,d) are the 5$^{th}$ and 10$^{th}$ ensembles, respectively (Table S1). Ensemble 5 has a mean aspect ratio of 3.7, with average nanorod lengths of 64.2nm and diameters of 17.9nm, while ensemble 10 has corresponding values of 7.7, 114.8nm, and 15.1nm. The mean and the standard deviation of the size and aspect ratio were obtained via image analysis of STEM images (Image J). A sampling population of 1000 nanorods was used. To validate the 1k sampling population, the results obtained from increasing sampling population were compared up to 10k. We found that increasing sampling size above 1000 did not significantly change the result (Fig.S1). Accurate estimation of the molar extinction coefficient depends crucially on the precise size measurement and quantification of the gold concentration, which determines the particle concentration. The gold concentration was obtained from ICP analysis as well as complementary estimation from 400 nm peak intensity (Fig.S2).

Alongside the STEM images are extinction, absorption, and scattering cross-sections in Figs. 2(a,c). The extinction spectrum is obtained using Cary 5000 spectrometer. The measured extinction intensity is converted to a cross-section by calculating the molecular extinction coefficient and normalizing it based on the volume of the particles according to Beer-Lambert law.[44] To measure the scattering and absorption contribution separately, we utilize a spectrophotometer with an integrating sphere detector[45,46] (a detailed descriptions of the measurements is provided in Fig. S3 and Table S1). It is assumed that the measured absorbance with the solution inside the integrating sphere detector is equal to the molecular absorption. The molecular scattering was obtained by subtracting molecular absorption from the extinction.

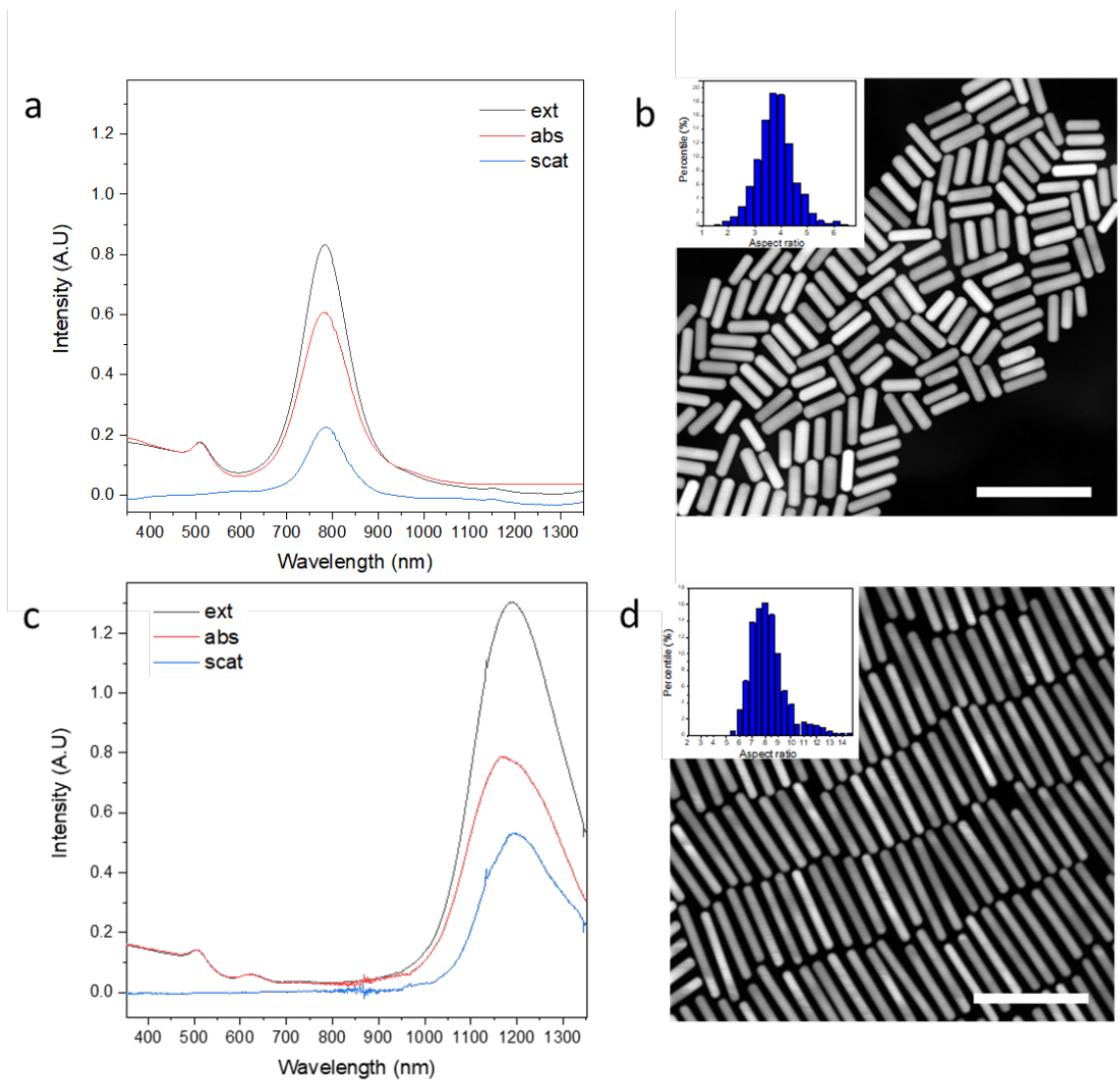

Figure 2. Representative results of optical and structural characterization of AuNRs. (a) Extinction, absorption and scattering cross-sections of AuNR ensemble 5 (mean aspect ratio of 3.7) and (b) a representative STEM image. The inset shows the distribution of aspect ratios. (c) Extinction, absorption and scattering cross-sections of AuNR ensemble 10 (mean aspect ratio of 7.7), alongside (d) A representative STEM image. For both STEM images the scale bar is 100 nm.

The shape of the extinction spectra of Fig. 2(a,c) reflect excellent shape purity and narrow polydispersity of the AuNR ensembles. One simple metric to evaluate this is by dividing the extinction intensity at the localized-surface-plasmon wavelength by the intensity at 400 nm wavelength, which is directly proportional to the concentration of gold atoms (Au (0)) and minimally influenced by structural resonances[47]: our shown ensembles achieve ratios of 5.2 and 9.1, which are significantly higher than the ratios obtained from conventional seed-mediated synthesis in the literature.[48] Large ratios imply less byproduct and lower polydispersity.[48,49] Note that there are a few synthetic protocols recently developed to synthesize AuNRs with comparable ratio via sophisticated modification of growth process.[50] The product quality is quantitatively confirmed by representative STEM images of the samples, as shown in Figs. 2(b,d). The shape purity is as low as 99%, while the variations in dimensions are small (Table S1). For a quantitative measure of the polydispersity, we select the relative standard deviation (RSD) of the aspect ratio, which is the standard deviation as a percentage of the mean. The ensembles shown have RSDs as small as 14% (Table S1). This percentage is a key single measure of the structural variations, as the aspect ratio is the primary determinant of the resonant wavelength (at highly subwavelength sizes), and variations in the aspect ratio are the key cause of the broadening of the scattering spectra. The high quality of the product can be ascribed to the new synthesis techniques described above.

**V. Polydispersity: Theory and Experiment**

For an ensemble of nanoparticles to collectively reach the bounds of Eqs. (2-4), the individual nanoparticles should all have identical resonant wavelengths and radiative efficiencies. According to the antenna model of Sec. III, this implies that nanorods should have identical aspect ratios and volumes. Yet in any experimental synthesis, as discussed above, invariably there is polydispersity in the volume and aspect ratio. (Variations of length and width are included in variations in volume and aspect ratio; since nanorods have only two geometric degrees of freedom, one can select any two of these four parameters as the independent degrees of freedom.) The key effect of polydispersity is that it creates a distribution of resonant wavelengths, with extinction cross-sections (per volume) that are reduced at the wavelength of interest. Thus, polydispersity broadens the collective spectral width and diminishes the peak.

For spherical AuNPs, typical synthetic methods generate particles with 10-20% RSD in diameter and this size dispersion is acceptable for common use.[51] However, wider polydispersity directly impacts performance in applications such as drug delivery[51,52], where the dimension itself is critical. Further purification is often necessary. It is more challenging to achieve small RSD in AuNR aspect ratio via typical solution-based synthetic approaches, which often contain subpopulations of particles of different shapes and sizes (e.g. spheres and cubes). Moreover, there is uncertainty in the measurements themselves, as the image analysis typically used to determine the means and distributions of the particle dimensions may be unrepresentative, due to limited sampling populations. The procedure for purification, sample preparation, and image capture/analysis can be lengthy. There have been limited efforts to estimate the mean and distribution of the nanoparticle dimension via other characterization tools to avoid biased measurement and to enable quick and robust measurements.[53-55]

The extinction spectrum of AuNRs has been used to quickly estimate some characteristics of colloidal AuNRs such as their aspect ratios, shape purity, and polydispersity.[48] For example, the nanoparticles' concentration can be derived from the absorbance at 400 nm.[47,56] The shape purity was accessed based on either the ratio of the magnitude of longitudinal-to-transverse surface plasmon resonance peak[49] or the ratio between L-LSPR to 400nm and the presence of a shoulder on the transverse band.[48] The full width at half maximum of the longitudinal LSPR band has been used as a measure of size dispersion.[57,58] However, most of the estimation is based on empirical databases, which can be inaccurate when attempting to account for many variables, including those listed above as well as the resonance energies and volumes of the rods.

Here, we seamlessly connect theory and experiment to provide a single extinction metric as an indicator of polydispersity. We describe the theoretical basis for our extinction metric and provide experimental confirmation of its utility. Taken together, our results further illuminate that polydispersity is the key constraint inhibiting the nanoparticles from reaching the bounds.

To model the effects of polydispersity across an ensemble, we start with the total extinction cross-section divided by the total nanoparticle volume, i.e. $\sigma_{\text{ext}}/V = (\sum_i \sigma_{\text{ext},i})/(\sum_i V_i)$, where $i$ indexes the individual nanoparticles in the ensemble. We can rewrite this expression in terms of the individual-nanoparticle per-volume cross-sections by multiplying and dividing the term in the numerator by $V_i$, i.e., $\sum_i \sigma_{\text{ext},i} = \sum_i V_i (\sigma_{\text{ext},i}/V_i)$. Thus the collective response can be written,

$$\frac{\sigma_{\text{ext}}}{V} = \frac{\sum_i V_i \left(\frac{\sigma_{\text{ext}}}{V}\right)_i}{V} = \sum_i f_i \left(\frac{\sigma_{\text{ext}}}{V}\right)_i, \tag{15}$$

where $f_i$ is the volume fraction of a particular nanoparticle, $f_i = V_i/V$. The volume fraction $f_i$ represents the percentage of nanoparticles with a particular volume and aspect ratio. Polydispersity enters in the term $(\sigma_{\text{ext}}/V)_i$, which is reduced at a particular wavelength due to shifts in the resonant wavelength.

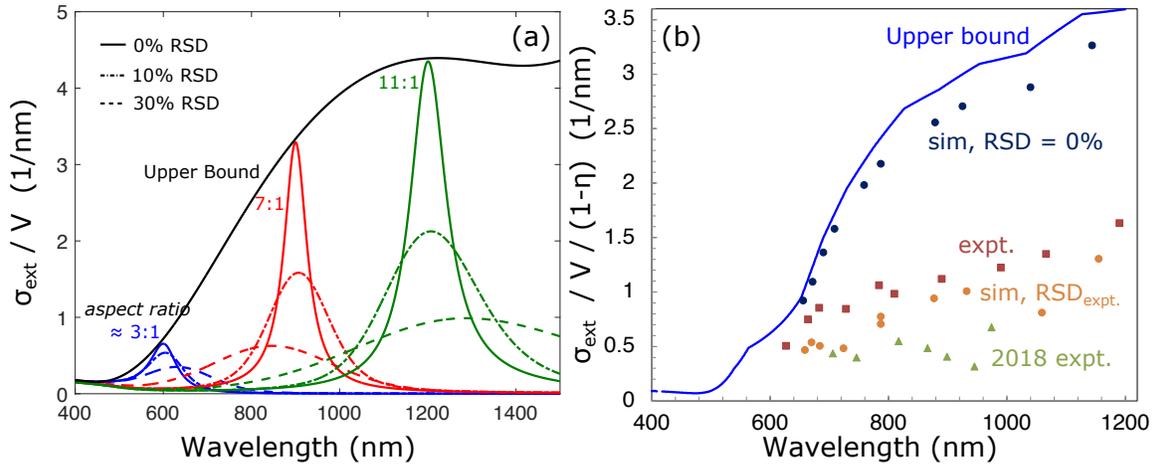

Figure 3 (a) Simulations demonstrating the effects of polydispersity, as measured by the nanorod aspect ratio relative standard deviation (RSD), on the magnitude of the optical response of a AuNR ensemble. Without any polydispersity (solid lines), nanorods with aspect ratios of approximately 3:1, 7:1, and 11:1 are able to reach the upper bound (solid black) for extinction cross-section per volume. The dot-dash and dashed lines are simulations with fixed volumes and Gaussian distributions of aspect ratios, with RSD at the 10% and 30% levels. (b) The theoretical upper bound to the extinction metric $\sigma_{\text{ext}}/V/(1-\eta)$ (solid line) is nearly approached by BEM simulations with RSD = 0% (blue circles). The ten experimentally synthesized and characterized AuNR ensembles (red) have measured RSD ranging from 14% to 23%; using the same RSD in BEM simulations leads to nearly matching theoretical predictions (orange circles). Experimental data from Ref. [28] (green triangles) shows a smaller extinction metric due to higher RSD. It is apparent that the synthesized nanorods are within factors of 1–2 of the bounds, and that the minimal polydispersity is the key factor responsible for the gap.

Figure 3a demonstrates the effect of polydispersity on the collective extinction per cross-section of an ensemble of plasmonic nanoparticles. To isolate the polydispersity effects we use a smoother Drude-Lorentz model of gold[59], but emphasize that the same effects are seen for tabulated experimental data, of gold or any other plasmonic material. The blue lines in Figure 3 are for quasistatic gold nanoparticles with average aspect ratios of 3.15:1 (to achieve resonant peaks at 600 nm wavelength). We consider polydispersity at the 0%, 10%, and 30% aspect-ratio RSD levels, and see significant inhibition of the peaks at the higher values. Similar computations are done for higher aspect ratios (7.1:1, red, peaking at 900 nm wavelength, and 10.87:1, green, peaking at 1200nm wavelength), for the same three levels of polydispersity. The upper bound of Eq. (2) is shown in the solid black line. Higher aspect ratios lead to longer-wavelength resonances, where the bound tends to be higher due to higher values of $|\chi|^2/\operatorname{Im}\chi$.

Our analysis further enables a means to experimentally characterize polydispersity from far-field scattering data in lieu of near-field imaging. We have identified two controlling parameters for the response of ellipsoidal plasmonic nanoparticles: aspect ratio, and volume relative to cubic wavelength. In tandem, these parameters control the resonant frequency and the radiative efficiency of the nanoparticles at that frequency. The extinction cross-section per volume is itself a useful measure of the total power the nanoparticles take from the incident waves, per unit material volume, but its maximum depends on the radiative efficiency of the underlying nanoparticles, which depends on their average aspect ratios as well as their volumes relative to the wavelength. But we can identify a new metric, which we term the "extinction metric," which is the extinction cross-section per volume divided by the $1-\eta$:

$$\text{Extinction metric} = \frac{\sigma_{\text{ext}}}{V(1-\eta)}. \qquad (16)$$

The extinction metric has two very useful properties: (1) on resonance it equals the quantity $(n_{\text{bg}}\omega/c)(|\chi|^2/\operatorname{Im}\chi)$, which depends only on the known material constants, and (2) it comprises easy-to-measure quantities: $\sigma_{\text{ext}}$ and $\eta$ can be computed from far-field scattering data, and $V$ can be measured via image analysis. In any experimental measurement of the extinction metric for nanoparticles, the extinction metric will invariably fall short of its on-resonance value (which also equals its upper bound) due to polydispersity. The degree to which the extinction metric falls short is a measure of how far off-resonance that subset of the nanoparticles is, which directly correlates with the polydispersity of the distribution. Thus the extinction metric enables quantitative identification of relative polydispersity with only far-field scattering data.

Figure 3b shows the extinction metric as a function of wavelength. The upper bound (now for tabulated gold optical constants[60]) is shown in solid blue and increases as a function of wavelength due to the increase in the normalized material metric, $(\omega/c)|\chi|^2/\operatorname{Im}\chi$, for gold in this range. The upper bound is compared to the experimentally obtained extinction metric for nanoparticle colloids with resonant wavelengths ranging from 627 nm to 1190 nm. The experimental extinction per volume shows a wavelength-dependence similar to that of the bound, but falls about a factor of 3 short in magnitude. We can theoretically show that this shortfall is primarily due to polydispersity (of both volume and aspect ratio). First, we perform BEM simulations of each colloid distribution using only the average nanoparticle lengths and widths, and assume 0% polydispersity. We model each nanoparticle as a circular cylinder with hemispherical endcaps, but the exact endcap shape simulated makes little difference. These simulations (blue circles) approach the upper bounds, as expected from the antenna model of Sec. III, and are similarly much larger than the experimental values. To model the synthesized colloids, we simulate their experimentally characterized length/width distributions. We take the individual length/width data collected via STEM, use a simple clustering algorithm to reduce the number of data points, and then run a BEM simulation for every remaining data point, for every colloid. These simulations, which account for polydispersity but no other non-ideality, result in the orange circles in Fig. 3(b), which closely track the experimental results. This confirms the claim that polydispersity is the key factor controlling how closely the upper bounds can be approach.

Also included in Fig. 3b are experimental measurements from Ref. [28]. It is apparent that the extinction metrics for the gold nanorods synthesized in this study are 1.5 to 2.6 times higher than that data, over the broad range of wavelengths. This is explained by the low polydispersity; as one indicator of this, the RSD of the aspect ratios range from 14% to 23% (Table S1), quite small values that are corroborated by the visual uniformity of the nanorods in the STEM images of Figs. 2(b,d). The improvements of these nanorods relative to previous results confirms the efficiency of the newly developed synthetic protocol.

Our theoretical framework can further provide quantitative estimates of polydispersity from the extinction metric, or vice versa. Figure 4 shows the dependence of the extinction metric as a function of polydispersity. There are two further variables to account for: the resonant wavelength, for which we include 700 nm (green), 800 nm (red), and 900 nm (blue) in the figure, as well as the average AuNR volume relative to a cubic wavelength. As expected intuitively, the extinction metric decreases as polydispersity increases, because higher polydispersity induces a broadening of the extinction spectrum. The degree to which polydispersity diminishes the extinction metric is also dependent on the average volume: larger volumes indicate larger nanoparticles, which are less sensitive to polydispersity as their spectra has already been broadened by their increased radiative efficiencies. Conversely, smaller nanoparticles by average volume see a more precipitous decline in their extinction metrics. Included in Fig. 4 are experimental data points (circular markers), using three linearly interpolated data points from the ten ensembles shown in Fig. 3(b). The average volumes of the ensembles range between $10^{-4}\lambda^3$ and $10^{-5}\lambda^3$, and lie just above the theoretical predictions. The close agreement of theory and experiment demonstrate the utility of this metric for nanoparticle characterization.

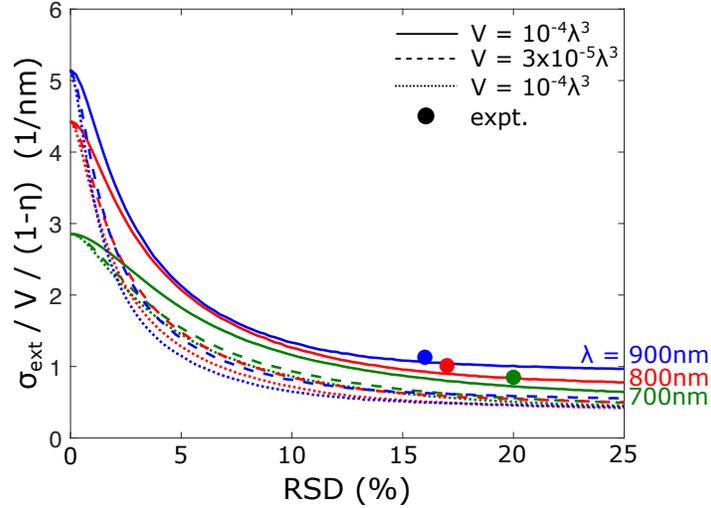

Figure 4. Extinction metric $\sigma_{\text{ext}}/V/(1-\eta)$ as a function of polydispersity. Increased polydispersity as measured by aspect-ratio RSD, reduces the extinction metric in a predictable and quantifiable way, for different wavelengths (blue, red, and green colors) as well as different average volumes (solid, dashed, and dotted lines) of the nanoparticles. Small-volume nanoparticles have smaller linewidths that make them more sensitive to polydispersity, while at longer wavelengths the value of $|\chi|^2/\operatorname{Im}\chi$ is larger, increasing the starting point for the extinction metric. The circular markers represent measured data from the AuNR ensembles, agreeing well with the theoretical predictions.

## VI. Conclusions

The key results of this work are three-fold: (1) in tandem, recent advances in experimental synthesis and theoretical bounds can enable colloidal nanorods to achieve maximum optical efficiencies, (2) polydispersity is a key variable to be minimized to achieve that maximum efficiency, and (3) polydispersity can be characterized through far-field scattering measurements encoded in a single "extinction metric." Looking forward, these results open multiple avenues of opportunity. After colloidal synthesis, post-processing of the nanoparticles is a mechanism for reducing polydispersity. The irradiation of gold nanorod colloids with a femtosecond laser can be tuned to induce controlled nanorod reshaping, yielding colloids with exceptionally narrow localized surface plasmon resonance bands (albeit at reduced solution volumes).[61] Per Figure 3, such reductions could enable 3X further enhancements of the peak optical response of gold nanoparticles, with even larger possible enhancements in materials such as silver or aluminum. From a materials perspective[37], the synthesis of lower-loss[62-64] and even possibly lossless[65] materials offers a pathway to even larger response than those of such conventional plasmonic materials, and could be combined with laser-reshaping and alternative polydispersity-reduction techniques. Another tantalizing opportunity is the synthesize of nanosized flakes of 2D materials, for which a similar optical-response bound framework can be applied[66], but whose underlying resonance theory (e.g. size and aspect-ratio dependencies) is very different due to the dimensionality. The potentially very small losses of graphene and similar materials[67] may enable larger efficiency enhancements, but they will require significant improvements in the controlled synthesis of such patterned 2D materials. Clearly, there is significant opportunity at the interface of theoretical design and chemical synthesis for maximum-efficiency optical nanoparticles.

# Supporting Information

# Towards Maximum Optical Efficiency of Ensembles of Colloidal Gold Nanorods

Owen Miller[1]*, Kyoungweon Park[2,3], Richard A. Vaia[2]*

1 Department of Applied Physics and Energy Sciences Institute, Yale University, New Haven, Connecticut 06511, USA

2 Materials and Manufacturing Directorate, Air Force Research Laboratory, Wright-Patterson AFB, Ohio 45433-7702, USA,

3 UES, Inc Dayton Ohio 45432, USA

**Table of Contents**

## SI 1.0 Methods

*Materials:*

Hexadecyltrimethylammonium bromide (CTAB) was purchased from GFS chemicals. $HAuCl_4$, $AgNO_3$, sodium borohydride and hydroquinone were purchased from Aldrich.

*Synthesis of Au NRs:*

The Au seeds were prepared according to the typical synthetic route.[1]  0.364g of CTAB was added to 10ml of 0.25 mM $HAuCl_4$. The solution was briefly sonicated (30 sec) and kept in a warm water bath (40°C) for 5 min to completely dissolve CTAB and left at 25°C for 10 min (solution A). A 0.01 M $NaBH_4$ solution was prepared and refrigerated (3°C) for 10min. 0.6ml of 0.01M $NaBH_4$ solution was quickly added drop wise to solution A while stirring at 800 rpm causing the color of solution to become light brown as Au seeds form. Stirring continued for 1 minute before aging the seeds for 5 minutes prior to use in all experiments.

The AuNRs were prepared according to the scale up protocol.[2,3] The growth solution was prepared by mixing $HAuCl_4$ (500 µL, 0.1 M), $AgNO_3$ (500 µL, 0.1 M), CTAB (0.1g). Next, hydroquinone (1.25 ml, 0.1M) was added to the growth solution as a mild reducing agent. Initially, 350 µl of growth solution was added into seed solution. After 2 hours, an aliquot of growth solution was added at certain intervals to obtain a targeted aspect ratio and volume of the rods. The number of addition, the volume of the aliquot, and the length of the interval determines the final aspect ratio and volume of the rods. The as-made solution was centrifuged at 3000 rpm for 20 min to remove large AgBr particles. The supernatant containing AuNRs was collected and centrifuged at 12,000 rpm for 20 min. Nine tenths of the supernatant was discarded and the concentrated AuNR sediment was collected and redispersed in 5 mM CTAB solution. The centrifugation was repeated 2 more times to ensure [CTAB] in AuNR dispersion to be 5 mM  The stock solution was diluted with 5 mM CTAB solution to obtain a proper optical intensity.

**Optical and TEM Characterization.** Ensemble extinction spectra were acquired with a Cary 5000 UV−vis−NIR spectrophotometer (Agilent) from 200 to 1350 nm. The absorption spectra were obtained by attaching diffuse reflectance accessories (integrating sphere) from 350 nm to

1350 nm. The scattering spectra were calculated by subtracting the absorption from the extinction for each sample measurement. Morphology and mean size of nanoparticles were determined by TEM and STEM (FEI Talos at 200 kV). For each sample, more than 1000 particles were measured to obtain the average size and the size distribution (Image J, NIST).

Absorbance measurements were performed with a Cary 5000 spectrometer equipped with a 150 mm integrating sphere (IS) detector. The spectrum was scanned from 200 nm to 1350 nm in increments of 1 nm with an integration time of 0.1 s, and a slit width of 2 nm.

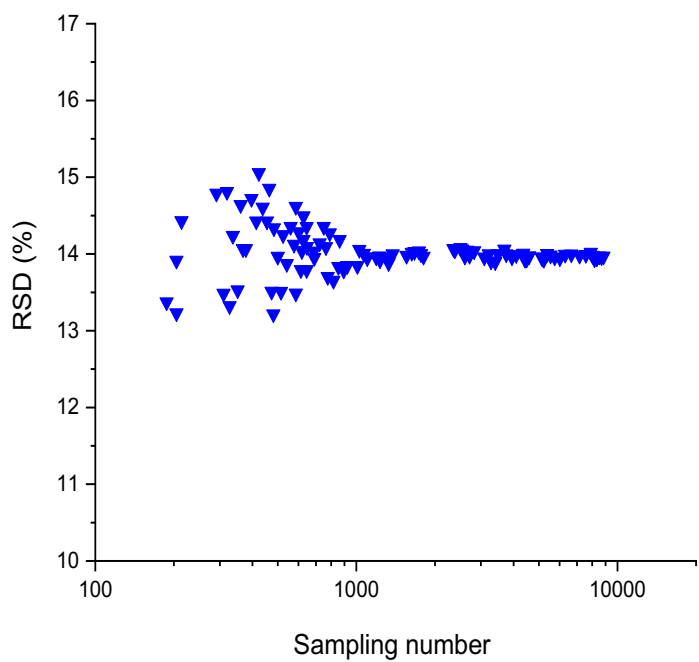

Fig. S1. The relationship between the sampling number and RSD (%) of aspect ratio (data obtained from NR 2 shown in Table S1). When the sampling number exceeds 1000, the RSD does not significantly change (within 0.2%).

**SI 2.0  Estimation of optical response from experimental measurement**

The molar extinction coefficients ($\varepsilon$) were experimentally determined using Beer−Lambert equation, $A = \varepsilon \cdot l \cdot c$, where A is the optical density of the solution, l is the light path length, and c is the AuNR concentration determined by the intensity at 400 nm which is validated by ICP-OES analysis. It has been known from the previous validation that A = 1.2 corresponds to [Au (0)] =

0.5 mM within a range of particle size.[1] We found that the previous empirical equation slightly underestimates gold concentration for the volume of the rods used in our experiment. The equation was adjusted for better estimation as shown in Fig. S1 where A = 1.07 corresponds to [Au (0)] = 0.5 mM.

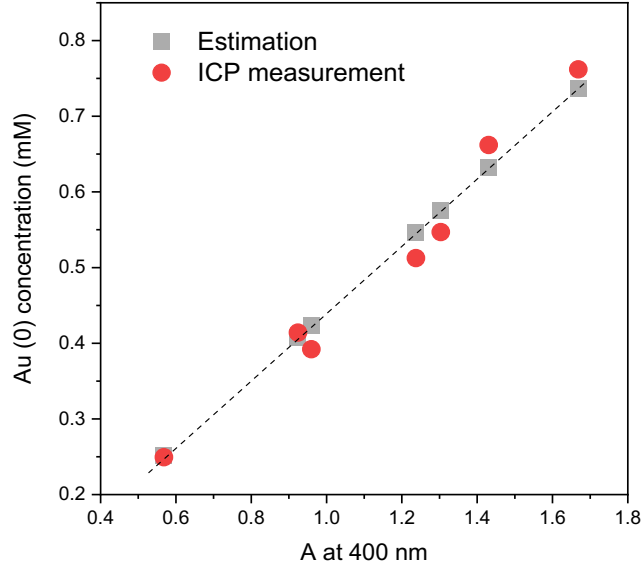

Fig. S2.  The relationship between the optical density (A) at 400 nm and Au (0) concentration. The adjusted empirical equation (grey dash line) is in good agreement with the measurement from ICP-OES analysis.

The molar extinction coefficient ε ( L mol$^{-1}$ cm$^{-1}$) is directly related to the extinction cross section σ (in units of cm$^2$) via the Avogadro constant:

$$\sigma = 1000 \ln(10) \frac{\varepsilon}{N_A} = 3.82 \times 10^{-21} \varepsilon$$

The calculation of extinction cross section of an isolated nanorod assumes that the incident light is linearly polarized along the long axis of the rod. Therefore, only longitudinal peak shows up. In the ensemble measurement, Au NRs are randomly oriented with respect to the propagation vector of the incident light. To compare the experimental values with theoretical value which is calculated (simulated) based on the propagation vector of the incident light being parallel to the axis of the rod, experimental values must be rescaled to account for the random orientation distribution of experimental ensemble.[4]   In the steady state, the nanorods rotate in all directions with equal probability. The orientation distribution of the AuNR major axis is then homogeneous. [5]

$$C_{ext,steady,z} = \frac{1}{4\pi} \int_0^{2\pi} \int_0^{\pi} C_{ext,(\theta,\varphi),z} \sin\theta \, d\theta \, d\varphi = \frac{1}{3} C_{ext,\theta=0}$$

$$C_{ext,\theta} = C_{ext,\theta=0} \cos^2\theta$$

Thus the factor of 1/3 was used to rescale experimentally obtained extinction cross section.

The extinction cross section was deconvoluted into scattering and absorption components by measuring the absorption cross-section with an integrating sphere. The scattering efficiency ($\eta$) is defined as the ratio of the scattering cross section ($\sigma_{scat}$) to the total extinction cross section ($\sigma_{ext}$) at the resonance and is expressed as:

$$\eta = \left|\frac{\sigma_{sca}}{\sigma_{ext}}\right|_{Res}$$

Extinction metric, $\sigma_{ext}/(1-\eta)V$ is calculated based on scattering efficiency measurement and the average volume estimated from image analysis.

## SI 3.0 Optical and Structural Characterization of AuNRs

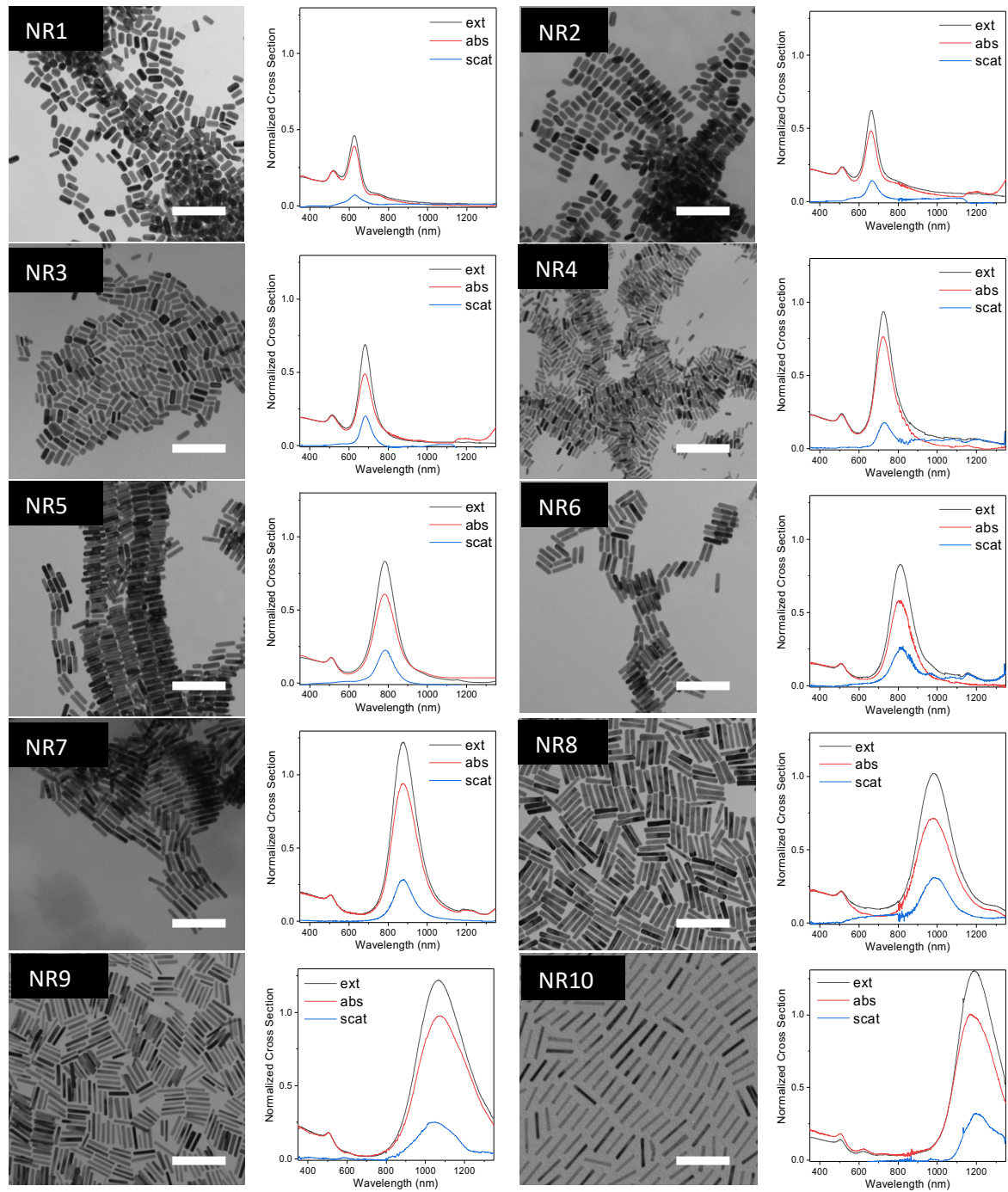

Fig. S 3. TEM images and UV-Vis-NIR spectra of the AuNRs. The scale bar is 200 nm.

Table 1. Summary of optical and physical characterization of AuNRs

| | L-LSPR (nm) | extinction cross section | Scattering efficiency | Aspect ratio | | Length (nm) | | Width (nm) | | Volume (nm³) | | Shape purity (%) |
|---|---|---|---|---|---|---|---|---|---|---|---|---|
| | | | | mean | stdev | mean | stdev | mean | stdev | mean | stdev | |
| NR1 | 627 | 1.07E+03 | 0.15 | 2.15 | 0.35 | 48.43 | 5.3 | 22.98 | 3.5 | 17424.93 | 5649.41 | 97 |
| NR2 | 664 | 1.02E+03 | 0.23 | 2.36 | 0.34 | 52.83 | 6.21 | 22.69 | 3.28 | 18876.23 | 6347.51 | 98 |
| NR3 | 683 | 8.35E+02 | 0.25 | 2.76 | 0.53 | 49.8 | 7.72 | 18.57 | 4.03 | 12779.4 | 6442.96 | 97 |
| NR4 | 728 | 3.29E+02 | 0.18 | 3.17 | 0.7 | 36.07 | 6.3 | 11.76 | 2.67 | 3772.34 | 2070.28 | 98 |
| NR5 | 784 | 4.94E+02 | 0.27 | 3.64 | 0.7 | 64.18 | 9.56 | 17.92 | 2.36 | 14977.56 | 4726.34 | 96 |
| NR6 | 810 | 7.08E+02 | 0.27 | 3.88 | 0.61 | 71.73 | 6.52 | 18.77 | 2.3 | 18323.91 | 4619.54 | 97 |
| NR7 | 890 | 6.47E+02 | 0.2 | 4.8 | 0.75 | 84.94 | 7.2 | 17.95 | 2.16 | 20174.14 | 4657.18 | 98 |
| NR8 | 990 | 9.09E+02 | 0.3 | 5.23 | 0.97 | 98.52 | 7.86 | 19.27 | 2.71 | 27236.9 | 7140.2 | 98 |
| NR9 | 1066 | 4.09E+02 | 0.2 | 6.71 | 1.55 | 83.58 | 12.48 | 12.93 | 2.82 | 11024.66 | 5099.62 | 99 |
| NR10 | 1190 | 3.38E+02 | 0.16 | 7.69 | 1.2 | 114.79 | 14.9 | 15.07 | 1.46 | 19775.65 | 5035.43 | 99 |

**SI 4.0 Radiative-Efficiency-Constrained Optical Cross-Section Bounds**

In this section we derive the bounds (fundamental limits) for cross-sections that are presented in the main text. We start with the expressions for the extinguished, absorbed, and scattered powers:

$$P_{\text{ext}} = \frac{\omega}{2} \text{Im} \int_V \bm{E}_{\text{inc}}^* \cdot \bm{P} \, \mathrm{d}x \tag{S1}$$

$$P_{\text{abs}} = \frac{\omega}{2} \frac{\text{Im}\,\chi}{|\chi|^2} \int_V |\bm{P}|^2 \, \mathrm{d}x \tag{S2}$$

$$P_{\text{scat}} = P_{\text{ext}} - P_{\text{abs}} \tag{S3}$$

We can now work out bounds on each of the three quantities, subject to the constraint on radiative efficiency, that $P_{\text{abs}} \leq (1-\eta)P_{\text{ext}}$. We start with extinction, whose bound is the solution of the optimization problem:

$$\max_{\bm{p}} \frac{\omega}{2} \text{Im}\left(\bm{e}_{\text{inc}}^\dagger \bm{p}\right)$$

$$\text{s.t.} \frac{\text{Im}\,\chi}{|\chi|^2} \bm{p}^\dagger \bm{p} \leq (1-\eta)\,\text{Im}(\bm{e}_{\text{inc}}^\dagger \bm{p}),$$

where to simplify notation we assume a discrete numerical basis, for which $e_{\text{inc}}$ and $p$ are discrete versions of their continuous counterparts, and the dagger symbol indicates the inner product, counterpart to spatial integration in the continuous case. For this optimization problem one can form the Lagrangian $L$, with Lagrange multiplier $\lambda$ (rescaled by $\omega/2$):

$$L = \frac{\omega}{2}\left[\text{Im}(e_{\text{inc}}^\dagger p) + \lambda\left(\frac{\text{Im}\,\chi}{|\chi|^2}p^\dagger p - (1-\eta)\,\text{Im}(e_{\text{inc}}^\dagger p)\right)\right].$$

Differentiating with respect to $p^\dagger$ (while keeping $p$ fixed) gives:

$$\frac{\partial L}{\partial p^\dagger} = \frac{\omega}{2}\left[\frac{i}{2}(1-\lambda(1-\eta))e_{\text{inc}} + \lambda\frac{\text{Im}\,\chi}{|\chi|^2}p\right]. \tag{S4}$$

Setting Eq. (4) equal to zero gives an expression for the optimal polarization field $p$; substituting that expression into the radiative-efficiency constraint yields the value of the Lagrange multiplier, $\lambda = -1/(1-\eta)$. Finally, the optimal $p$ is given by $p = i(1-\eta)(|\chi|^2/\text{Im}\,\chi)e_{\text{inc}}$, giving a bound on the extinction power of $P_{\text{ext}} \leq \frac{\varepsilon_0 \omega}{2}(1-\eta)|E_0|^2 V$, where we have taken an incident plane wave with amplitude $E_0$, a scatterer volume $V$, and re-introduced the free-space permittivity $\varepsilon_0$ that we had dropped for convenience earlier. Then, the cross-section is given by the extinguished power divided by the incident plane-wave intensity, which is given by $I_{\text{inc}} = \frac{1}{2Z_{\text{bg}}}|E_0|^2$. For a nonmagnetic medium, the background impedance is given by $Z_{\text{bg}} = n_{\text{bg}}Z_0$, where $n_{\text{bg}}$ is the background refractive index. Finally, inserting all of these expressions leads to the bound on cross-section per volume of (replacing $\varepsilon_0/Z_0$ with $c$, the speed of light)

$$\frac{\sigma_{\text{ext}}}{V} \leq (1-\eta)\frac{n_{\text{bg}}\omega}{c}\frac{|\chi|^2}{\text{Im}\,\chi}. \tag{S5}$$

We can repeat this process with the other two power quantities, absorption and scattering. The process for absorption is very similar. Our optimization problem is now

$$\max_{\mathbf{p}} \frac{\omega}{2} \frac{\operatorname{Im} \chi}{|\chi|^2} \mathbf{p}^\dagger \mathbf{p}$$

$$\text{s.t.} \frac{\operatorname{Im} \chi}{|\chi|^2} \mathbf{p}^\dagger \mathbf{p} \leq (1-\eta) \operatorname{Im}(\mathbf{e}_{\text{inc}}^\dagger \mathbf{p}),$$

leading to a Lagrangian of the form

$$L = \frac{\omega}{2}\left[\frac{\operatorname{Im}\chi}{|\chi|^2}\mathbf{p}^\dagger\mathbf{p} + \lambda\left(\frac{\operatorname{Im}\chi}{|\chi|^2}\mathbf{p}^\dagger\mathbf{p} - (1-\eta)\operatorname{Im}(\mathbf{e}_{\text{inc}}^\dagger\mathbf{p})\right)\right].$$

Now, differentiation and substitution lead to the same optimal $\mathbf{p}$ of $\mathbf{p} = i(1-\eta)\mathbf{e}_{\text{inc}}$, leading to an absorption bound of $P_{\text{abs}} \leq \frac{\varepsilon_0 \omega}{2}\left(\frac{|\chi|^2}{\operatorname{Im}\chi}\right)(1-\eta)^2 |E_0|^2 V$. Dividing by the incident intensity of the plane wave, the cross-section bound is:

$$\frac{\sigma_{\text{abs}}}{V} \leq (1-\eta)^2 \frac{n_{\text{bg}}\omega}{c} \frac{|\chi|^2}{\operatorname{Im}\chi}. \tag{S6}$$

Finally, the scattered power can be optimized in the same way. It turns out the optimal currents are the same, for the third, time: $\mathbf{p} = i(1-\eta)\mathbf{e}_{\text{inc}}$, which leads to a cross-section bound of

$$\frac{\sigma_{\text{scat}}}{V} \leq \eta(1-\eta)\frac{n_{\text{bg}}\omega}{c}\frac{|\chi|^2}{\operatorname{Im}\chi}. \tag{S7}$$

Thus, we have derived the three key bounds of the manuscript.

## SI 5.0 Antenna-Model-Based Expressions for Nanorod Cross-Sections

We start with the expression for the radiative efficiency given in the main text:

$$\eta = \frac{V/\lambda^3}{V/\lambda^3 + \frac{3}{4\pi^2}\frac{\mathrm{Im}\,\chi}{|\chi|^2}}.$$

We can equivalently write this in terms of the volume of the scatterer, finding

$$V/\lambda^3 = \frac{\eta}{1-\eta}\frac{3}{4\pi^2}\frac{\mathrm{Im}\,\chi}{|\chi|^2}.$$

Taking this one step further, we can even use it to write the cube of the resonant wavelength in terms of radiative efficiency, volume, and material loss:

$$\lambda^3 = \frac{1-\eta}{\eta}\frac{4\pi^2}{3}\frac{|\chi|^2}{\mathrm{Im}\,\chi}V.$$

Then we can write the extinction cross-section of the antenna, for example, as

$$\sigma_{\mathrm{ext}} = \eta\frac{3\lambda^2}{2\pi} = \frac{3\eta}{2\pi\lambda}\lambda^3 = \frac{3\eta}{2\pi\lambda}\frac{1-\eta}{\eta}\frac{4\pi^2}{3}\frac{|\chi|^2}{\mathrm{Im}\,\chi}V = \frac{2\pi}{\lambda}(1-\eta)\frac{|\chi|^2}{\mathrm{Im}\,\chi}V = \frac{n_{\mathrm{bg}}\omega}{c}(1-\eta)\frac{|\chi|^2}{\mathrm{Im}\,\chi}V.$$

This is exactly the expression given in the main text, and matches the bound expression as well.

The derivations for the absorbed and scattered powers follow exactly the same procedure.